\documentclass[pre,aps,twocolumn,showpacs,floatfix,eqsecnum]{revtex4}
\usepackage{graphicx}

\begin{document}

\def\OMIT#1 {}
\def\LATER#1 {}
\newcommand{\eqr}[1]{(\ref{#1})}
\newcommand{\half}{{\textstyle{1\over2}}}
\newcommand{\fourth}{{\textstyle{1\over4}}}

\def \kag {kagom\'e }
\def \pb {{\overline {\bf p}}}
\def \n {{\hat {\bf n}}}
\def \p {{\bf p}}
\def \hh {{\bf h}}
\def \qq {{\bf q}}
\def \ss {{\bf s}}
\def \G {{\bf G}}
\def \uu {{\bf u}}
\def \pp {{\bf p}}
\def \Pol {{\bf P}}
\def \QQ {{\bf Q}}
\def \RR {{\bf R}}
\def \rr {{\bf r}}
\def\JJ{{\bf J}}
\def \kk {{\bf k}}
\def \KK {{\bf K}}
\def\LL{{\bf L}}
\def \HH {{\cal H}}
\def \F {{\bf F}}
\def\Hfield{{\bf H}}
\def\Jspin{{J_{\rm spin}}}
\newcommand{\la}{\langle}
\newcommand{\ra}{\rangle}
\def\Struc#1{{\la |\tt(#1) |^2\ra}}
\def \tt {{\tilde t}}
\def \tPol {{\tilde {\Pol}}}
\def\Tcmf{{T_c^{\rm MF}}}
\def\betaeff{{\beta_{\rm eff}}}

\title{POWER-LAW SPIN CORRELATIONS 
IN PYROCHLORE ANTIFERROMAGNETS}

\author{C.~L.~Henley}

\affiliation{Department of Physics, 
Cornell University, Ithaca, New York 14853-2501}

\begin{abstract}
The ground state ensemble of 
the highly frustrated pyrochlore-lattice antiferromagnet
can be mapped to a coarse-grained ``polarization''
field satisfying a zero-divergence  condition 
From this it follows that the correlations of this field, as
well as the actual spin correlations, decay with separation 
like a dipole-dipole interaction ($1/|R|^3$).
Furthermore, a lattice version of the derivation gives an 
approximate formula for spin correlations, with several
features  that agree well with 
simulations and neutron-diffraction measurements of diffuse scattering, 
in particular the pinch-point (pseudo-dipolar) singularities 
at  reciprocal lattice vectors. 
This system is compared to others in which constraints also imply
diffraction singularities, and other possible applications of the
coarse-grained polarization  are discussed.
\end{abstract}

\maketitle

\section{introduction}
\label{sec:intro}

Highly frustrated antiferromagnets are characterized by
a very large number of essentially degenerate states,
such that in a range of temperatures much smaller than the 
spin interaction scale 
they have strong local correlations,  yet fail to order~\cite{ramirez}:
diffuse scattering is the obvious probe of such a state.
In this paper, I make the point that the classical ground-state
ensemble can entail constraints that ensure a generic
power-law decay of correlations: a picture of theses state 
as ``liquid-like'' (i.e. featureless) is thus imcomplete.


We consider specifically the pyrochlore lattice, 
consisting of corner-sharing tetrahedra, partly
because of its simplicity as a model
[high (cubic) symmetry and nearest-neighbor interactions], 
but most importantly because it is the magnetic
lattice  in many real systems:
the B-site spinels~\cite{lee00,lee02},
two families of pyrochlores represented by 
$\rm CsNiCrF_6$~\cite{harris94,harris97a}, 
and the oxides $\rm Y_2Mn_2O_7$~\cite{Y2Mo2O7}, and
even the transition-metal lattice in ``frustrated 
itinerant'' antiferro- or ferrimagnetic Laves phases, such as
$\rm Y_{1-x}Sc_xMn_2$~\cite{shi93,ballou}. 
Ferroelectric degrees of freedom (water ice~\cite{bernal-pauling,youngblood})
and charge order (in magnetite~\cite{an56}) are also equivalent
to pyrochlore systems.

Different pyrochlores  show a wide variety of magnetic behaviors
at low temperatures:
apart from long-range order, these include 
lattice distortions~\cite{lee00,tcherny}, 
``spin-ice'' behavior~\cite{harris97b,ramirez-spinice,Si99,Bra01a,Bra01b},
paramagnetism (in $\rm Tb_2Ti_2O_7$)
down to low temperatures,~\cite{Tb2Ti2O7}
or spin glass freezing (in $\rm Y_2Mo_2O_7$
even when structural disorder is quite small~\cite{Y2Mo2O7}.
However, most of these have a cooperative paramagnet regime
at higher temperatures -- but still quite low compared to
the interaction scale $J$ -- demonstrating that they have
a large set of nearly degenerate (and accessible) states.
The present paper addresses this phase, as it would be
extrapolated to zero temperature.

In the remainder of this section, I review the lattice and
the ground states of a pyrochlore antiferromagnet, 
and survey three situations in which it can
be modeled by the ensembles used in this paper.
Then (Sec.~\ref{sec:coarse}
the mapping of the low-temperature states
of the Ising model to the ``diamond ice'' model
is to construct a coarse-grained polarization field,
which functions somewhat like an order parameter
for this model, and this is used to
predict some unusual features of the magnetic  diffuse scattering.
Sec.~\ref{sec:lattice} presents another version of the
derivation, which not only gets the long-wavelength behavior
(corresponding to the neighborhoods of special points in the 
Brillouin zone) but provides an approximation for the entire zone. 
I also discuss (Sec.~\ref{sec:disc}) other situations in
which local constraints in a ground state produce diffraction
singularities, and speculate on other questions that can
be addressed using the insight that the polarization field
contains the important long-wavelength degrees of freedom.

A parallel paper~\cite{moessner-dipolar} contains many of the same results.

\subsection{Pyrochlore lattice and Hamiltonian}

\begin{figure}[ht]
\includegraphics[width=1.0\linewidth]{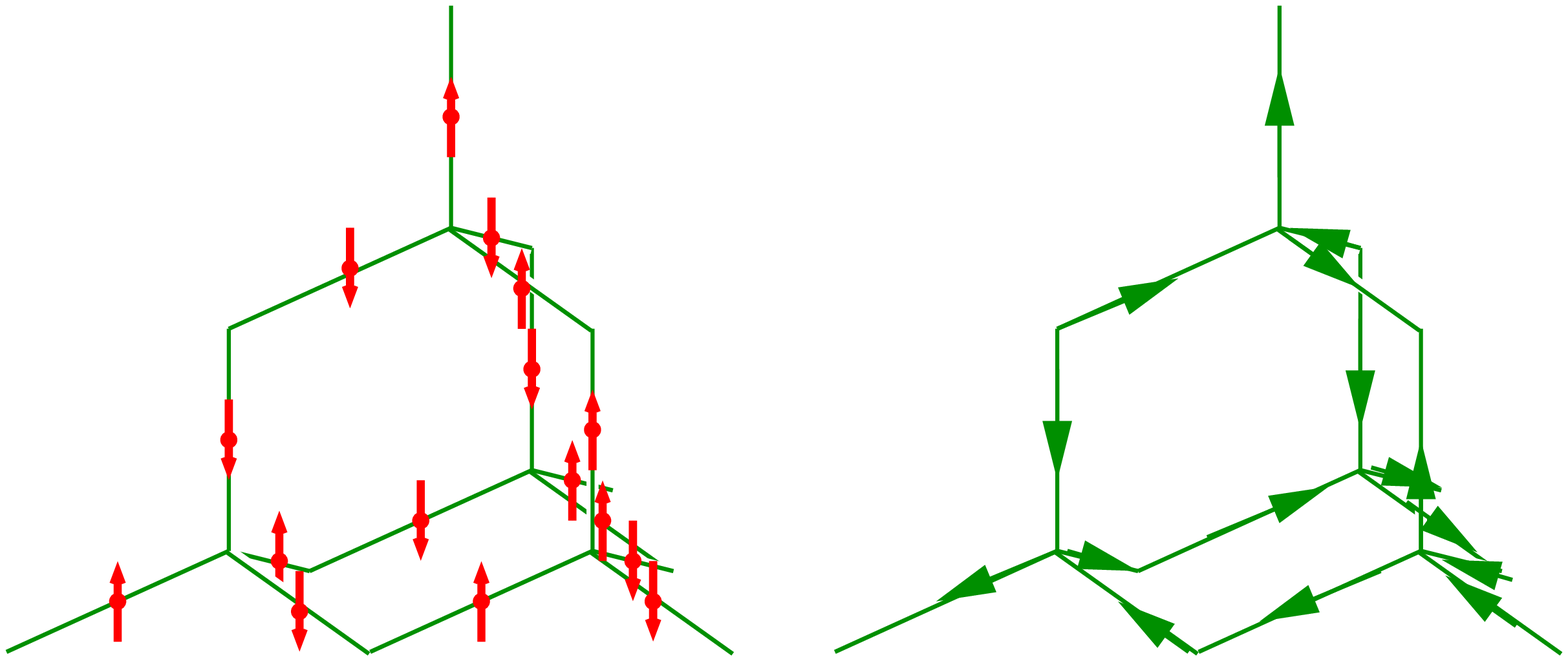}
\caption{
\LATER{NEED TO REDRAFT FIGURE TO MAKE IT VISUALLY CLEAR.}
(a) Ising ground state on fragment of pyrochlore lattice;
spins are shown on edges of a diamond lattice.
(b). The corrsponding ice-model arrows between the diamond vertices.}
\label{fig:lattice}
\end{figure}

The pyrochlore lattice consists of corner-sharing tetrahedra
arranged in cubic (fcc) symmetry.
Its key property is that
the sites are the bond midpoints of a diamond lattice.
[See Fig.~\ref{fig:lattice};
other figures emphasizing the tetrahedra are in
Refs.~\onlinecite{tcherny} or~\onlinecite{moessner98a,moessner98b}.]
For future use, we define vectors pointing towards the four corners 
of a tetrahedron, 
\LATER{Use ``cases''? or eqalign?}
 \begin{eqnarray}
     \ \uu_{1,2,3} \equiv [\fourth,-\fourth,-\fourth] \quad
      \nonumber \\
     \uu_4 \equiv [\fourth, \fourth, \fourth]  \qquad
     + \text{cyclic perms}.
 \label{eq:uu}
 \end{eqnarray}
In the diamond lattice, every even-to-odd bond vector is a $\uu_m$.  

The basic results derived in this paper are transferable to other
lattices in which the spins sit on the bonds of a bipartite
lattice, and with antiferromagnetic interactions among
all spins on bonds sharing a 
common endpoint,such that their sum is constrained
to a fixed value.
Besides the pyrochlore lattice,
this class includes the kagom\'e lattice, 
the two-dimensional
``checkerboard'' lattice,~\cite{moessner98a,moessner98b},
the ``sandwich'' lattice modeling SCGO and consisting 
of two kagom\'e layers linked by an additional 
triangular layer of 
spins,~\cite{broholm-SCGO,henley-HFM00,kawamura}
and the garnet lattice~\cite{petrenko}.

The spin Hamiltonian, 
having only nearest-neighbor antiferromagnetic exchange,
can be cast as a sum of squares:
   \begin{equation}
       \HH_{\rm spin}= \Jspin 
        \sum _{\la ij \ra} \ss_i \cdot \ss_j +{\rm const}
       = \frac{\Jspin}{2} \sum _\alpha \LL_\alpha^2, 
   \label{eq:Hspin}
   \end{equation}
where the total spin of four sites in a tetrahedron is
   \begin{equation}
      \LL_\alpha \equiv \sum _{i\in \alpha} \ss_i
   \label{eq:L}
   \end{equation}
where $\alpha$ labels each diamond site
and ``$i\in\alpha$'' runs over the surrounding four spins.
For simplicity, I will mostly treat the case with Ising spins $t_i$.
(There is little loss of generality, in view  of Sec.~\ref{sec:Escales}, below.)
   \begin{equation}
       \HH = \half J \sum _\alpha L_\alpha^2 
       = J \sum _{\la ij \ra} t_i t_j +{\rm const} .
   \label{eq:Ham}
   \end{equation}

From \eqr{eq:Ham} it is evident that any  state with 
   \begin{equation}
   L_\alpha = 0.
   \label{eq:ground}
   \end{equation}
is a classical ground state; these are massively degenerate.
The frustration of the pyrochlore  Ising model,
and the extensive ground state entropy implied by \eqr{eq:ground}, 
were recognized very early~\cite{an56}. 

\subsection{Temperature regimes and validity of Ising model}
\label{sec:Escales}

Here I discuss the three different situations in which realistic
models can be treated under this paper's scheme.
[Readers who are content with a treatment of an Ising toy model, 
and do not immediately demand any connection to experiment, 
may skip to Sec.~\ref{sec:coarse} where the central result
is derived.]

First, it is possible that spin anisotropies 
(additional to \eqr{eq:Hspin}) reduce the ground state
manifold to the Ising states.
If the lattice is to remain cubic, the preferred axis cannot
be the same for each spin, but instead is $\hat \uu_{m(i)}$, 
where $\hat \uu_m = 4 \uu_m /\sqrt{3}$ is the unit vector
of \eqr{eq:uu},
and $m(i)$ labels the direction of the diamond-lattice
bond on which spin $i$ sits.
Assuming the spins are strongly aligned, one puts
$\ss_i  = t_i \hat \uu_{m(i)}$ and the Hamiltonian
reduces to \eqr{eq:Ham} with $J=- \Jspin/3$, 
since $\hat  \uu_m \cdot \hat \uu_{m'} = -1/3$ for 
$m\neq m'$, and $m(i)\neq m(j)$ for nearest neighbors.
This is the ``spin-ice'' model~\cite{harris97b,ramirez-spinice}.

The other two situations correspond to isotropic Heisenberg 
models in the large-$S$ semiclassical regime (where $S$ is the
spin length).  This is necessary to guarantee that \eqr{eq:ground} is
a good starting point for specifying the ground states.
One can identify a succession of energy scales in 
the pure nearest-neighbor Heisenberg antiferromagnet, 
corresonding to successive breakings of the grond state
degeneracy as more effects are considered.

The largest scale is $E_J \sim zJS^2$  where $z$ is the
coordination number; this is the scale of the mean-field
(classcial) part of the exchange energy,  and of the Curie-Weiss
constant. At temperatures $T \ll E_J$, it is reasonable
to approximate the spin ensemble by a subset of the 
ground-state ensemble.

The spin-wave spectrum $\{\omega_j \}$ depends 
upon which state we linearized around.
Thus, the total spin-wave zero-point energy (to harmonic order, 
$\sum_j \half \hbar \omega_j$) 
partially breaks the ground-state degeneracy, favoring {\it collinear} states,
~\cite{shen82,hen89}. This is expressed quantitatively by an
effective Hamiltonian~\cite{oja,larhen90,henley-HFM00}, 
with coupling constant $\sim E_{\rm coll}$, where
the next largest scale is $E_{\rm coll} \sim JS \ll E_J$.
(Even when $S$  is not so large, e.g. $S=5/2$ as with real spins, 
the coefficient of $JS$ is commonly less than $1/10$ 
so the inequality is valid.)
In lattices other than pyrochlore, one finds a similar regime in
which a discrete subset of the classical ground states 
(e.g. coplanar states,~\cite{ritchey,chubukov} in the \kag case) gets
selected by the harmonic zero-point energy.

Finer treatment of the spin-wave fluctuations produces an 
even smaller energy scale $E_{\rm disc}$ for 
selection among the discrete states.  
In the \kag case,~\cite{chubukov,chan};
 this is due to anharmonic terms
in the Holstein-Primakoff expansion, 
so that $E_{\rm disc} \propto S^{2/3} \ll S$.
In the pyrochlore case, besides the anharmonic terms, 
$E_{\rm disc}$ does include contributions of $O(JS)$
from the harmonic terms.~\cite{pyro-Heff}.
However, these terms only partly 
break the degeneracy, and the coefficients are small, so
we can still assume $E_{\rm disc} \ll E_{\rm  coll}$.

Thus, the second regime in which the Ising theory
models the correlations is 
$E_{\rm disc} < T < E_{\rm coll}$.
One expects a symmetry breaking to
long-range collinear order, in which 
a global axis $\n$ is spontaneously adopted
and each spin is given by $\ss_i = t_i \n$,  with $t_i=\pm1$, 
plus small fluctuations.
This phase is a kind of ``spin nematic''~\cite{chancole91}.
The energy terms distinguishing different Ising states
are unimportant since $T> E_{\rm disc}$, hence
the spin ensemble is roughly the ground states of \eqr{eq:Ham}, 
as claimed.  
\OMIT{The phase of the XY (2-component) model, which 
develops long-range collinear order, 
would be analogous.~\cite{moessner-dipolar}}

Of course, in the real world there is another energy
scale $E_{\rm pert}$ for perturbations of the pure Heisenberg
Hamiltonian, representing the magnitude further-neighbor exchange and
dipole couplings, as well as lattice distortions,  and
disorder, which cause a transition or freezing into 
some other state at sufficiently low  temperature, so
the regime of the spin nematic-like  phase is really 
${\rm max}(E_{\rm disc},E_{\rm pert}) < T < E_{\rm coll}$.

\OMIT {Such a collinear ordering was apparently not observed in 
Monte Carlo simulations of the
{\it classical} Heisenberg system~\cite{reimers92a}}

Finally, the third situation is when $E_{\rm coll} \ll T \ll E_J$.
This ensemble is well modeled by the low-$T$ limit
of the classical Heisenberg model.
\OMIT{Note that, although thermal fluctuations
tend to favor local collinearity, in the Heisenberg spin 
case this does not diverge~\cite{moessner98a,moessner98b}
as $T\to 0$,and is insufficient to induce long-range order
of the collinear axis.}
It turns out  (Sec.~\ref{sec:Heis-pol}, below)
that the key notion of ``polarization'' does extend to the 
Heisenberg spin case.  I have mostly neglected that case in this
paper because, at my level of approximation, the results
look identical; it would merely obscure the notation by
adding spin-component indices everywhere.

\section{Coarse-graining, Fourier mode fluctuations,  
and long-range correlations}
\label{sec:coarse}

In this section, I present the steps leading to power-law 
correlations using the framework of a continuum theory, 
in the ideas appear more transparently.

\subsection{Ice mapping and local polarization}
\label{sec:icemap}

In fact the pyrochlore  (Ising) ground states 
map 1-to-1 onto those of the well-known diamond-lattice 
ice model~\cite{an56,liebmann86}, in which   the degrees of freedom are arrows
along the lattice bonds. 
In the map, every tetrahedron center becomes a vertex of 
the diamond lattice, while the spin sites map to bond centers 
of the diamond lattice.  Each spin $t_i =+ 1(-1)$ maps to an arrow 
pointing along the corresponding diamond lattice edge, 
in the positive (negative) sense from the even to the odd vertex.
This well-known mapping ~\cite{an56,liebmann86,Hen92} 
is also used to model the ``spin ice'' system Dy$_2$Ti$_2$O$_7$ 
(and also Ho$_2$Ti$_2$O$_7$), wherein 
local $\la 111 \ra$  anisotropy plus
{\it ferromagnetism} makes a highly frustrated 
Ising model~\cite{harris97b,ramirez-spinice,Si99,Bra01a}.)
The ground state condition -- net spin of every tetrahedron is zero 
-- maps to the ``ice rule'' 
the numbers of incoming and outgoing arrows are equal at every vertex
~\cite{bernal-pauling}.

The key object in this paper is the ice polarization field.
On a diamond-lattice bond a polarization 
$t_i \uu_{m(i)}$ can be defined, aligned from the even to odd diamond-lattice
vertex if the spin is up, oppositely if it is down;
here $m(i)$ is the local 3-fold axis of site $i$.
On every diamond vertex, we define the mean of this polarization 
over the surrounding tetrahedron of spins:
   \begin{equation}
        \Pol(\RR_\alpha) \equiv \sum _{i \in \alpha} 
        t_i \uu_{m(i)}.
   \label{eq:Pol-alpha}
   \end{equation}
The six possible ground-state configurations
of that tetrahedron correspond
to  $\Pol(\RR_\alpha)= (\pm 1,0,0)$, $(0,\pm 1, 0),$ or $(0,0,\pm 1)$.

Finally, the coarse-grained arrow field $\Pol(\rr)$ is the 
coarse-grained polarization averaged over some larger neighborhood and
assumed to vary smoothly.

\subsection{Effective free energy and correlations}
\label{eqc:FtotFT}

The ground-state entropy density is a function of the 
average polarization.  
A subensemble of states in which $\Pol$ is large 
(which can be forced by boundary conditions)
has relatively little freedom for rearrangements of the spins
or arrows; indeed, for a saturated polarization
such as $\Pol = (1,0,0)$ the ensemble consists
of a single microstate.
Thus it is very plausible that the entropy density has a 
maximum for zero polarization. 
Therefore, to lowest order, 
the total free energy (arising entirely from entropy), as
a function of coarse-grained $\Pol(\rr)$, has the form
   \begin{equation}
       F_{\rm tot} (\{ \Pol (\rr) \})/T = v_{\rm cell}^{-1}
  \int d^3\rr \half \kappa |\Pol(\rr)|^2, 
   \label{eq:Fcoarse}
   \end{equation}
where $v_{\rm cell}=a^3/4$ is the volume of a primitive unit cell.
The ``stiffness'' $\kappa$ is dimensionless,
as appropriate since it is purely entropic in origin. 

Corresponding to the condition \eqr{eq:ground}, i.e. the ice rule, 
$\Pol(\rr)$ satisfies  a divergence constraint
   \begin{equation}
       \nabla \cdot \Pol(\rr)=0
   \label{eq:divergence}
   \end{equation}
like a magnetic field without monopoles.
Eqs.~\eqr{eq:Fcoarse} and \eqr{eq:divergence} look, respectively,
like the field energy of a magnetic (or electric) field, 
and its divergence constraint,  in the absence of monopoles (or charges). 
These equations, together,
signify that that the probability distribution
of the (long-wavelength portion of) the polarization field is 
the (constrained) Gaussian distribution
   \begin{equation}
      {\rm Prob}(\{ \Pol(\rr) \} ) \propto
      e^{- F(\{ \Pol (\rr) \})/T}
      \prod _\rr \delta (\nabla \cdot \Pol (\rr))
   \label{eq:Poldist}
   \end{equation}

Fourier transforming \eqr{eq:Fcoarse} simply gives
$F_{\rm tot}= \sum _\kk \half \kappa |\Pol(\kk)|^2$, 
so a naive use of equipartition would give 
$\langle P_\mu (-\kk) P_\nu(\kk) \rangle  = (1/{\kappa}) \delta_{\mu\nu}$.
But the divergence constraint \eqr{eq:divergence} imposes
    \begin{equation}
         \kk\cdot  \Pol(\kk)=0
    \label{eq:div-Fourier}
    \end{equation}
in Fourier space. Thus the correct result has
the longitudinal fluctuations projected out:
   \begin{equation}
    \la  P_\mu(-\kk)  P _\nu(\kk)\ra 
    = \frac{1}{\kappa}\Bigl( \delta _{\mu\nu}  - 
     \frac {k_\mu k_\nu}{|\kk|^2} \Bigr),
   \label{eq:flucts}
   \end{equation}
Fourier transforming \eqr{eq:flucts} back to direct space gives
   \begin{equation}
       \langle P_\mu (0) P_\nu(\rr)\rangle \cong 
       \frac {4\pi}{\kappa} \Bigl[ \delta(\rr) + 
       \frac{1}{r^3} ( \delta _{\mu\nu} -3 \hat r_\mu\hat r_\nu) \Bigr]
   \label{eq:ice-dipolar}
   \end{equation}
at large separations $\rr$ (where $\hat \rr \equiv \rr/|\rr|.$)
{\it Correlations have  the
spatial dependence of a dipole-dipole interaction,}
which is a power law.
Models that exhibit such correlations
-- including the pyrochlore system,
so often described as ``liquid-like'', 
-- are thus, in a sense,  in a {\it critical} state.
[The generalization of \eqr{eq:ice-dipolar}
for $d$-dimensional real space.
would be a $1/ r^{d}$  decay.]

This  criticality was first appreciated in the ice model itself, 
being detected originally in a simulation~\cite{stillinger}.
A functional form with a dipolar singularity 
like \eqr{eq:flucts} was produced by 
a clever random-walk approximation to a series expansion~\cite{villain}.
The universal explanation, made here,  that dipolar correlations 
arise from \eqr{eq:Fcoarse} with the divergence condition,
was first put forward to explain experiments on
two-dimensional ice-like systems~\cite{youngblood}.

Ref.~\onlinecite{huse-dimer} 
have also presented an  ansatz equivalent to \eqr{eq:Poldist}, 
derived the dipolar correlations, 
and confirmed them by simulations, for
the dimer covering of a simple cubic lattice.
[See their eq.~(3).]
The $1/|\rr|^3$ decay has also been obtained analytically and
numerically in Ref.~\onlinecite{hermele} , for the 
pyrochlore model of this present paper.

\begin{figure}[ht]
\includegraphics[width=1.6\linewidth]{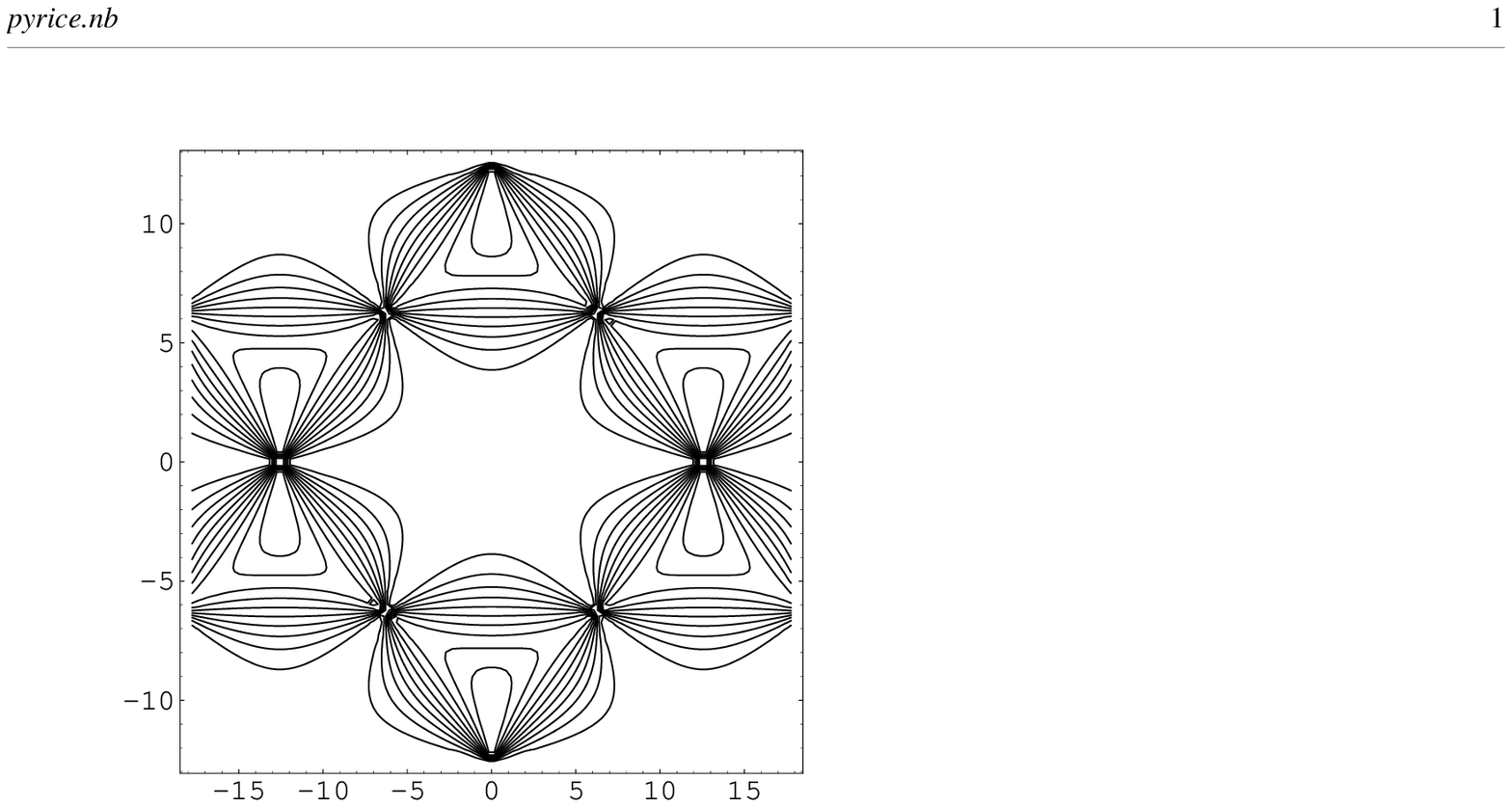}
\caption{
\LATER{CHECK I USED THE NEW, NOT 1995 FORMULA.}
Calculated structure factor in the plane normal to $[1 \bar 1 0]$.
(The figure is the same slice as shown in 
Refs.~\protect\onlinecite{zinhar95} and 
\protect\onlinecite{harris95}.)}
\label{fig:strucf}
\end{figure}

Ref.~\onlinecite{moessner98b}, Sec.~II~D 1, recognized
that the ground-state constraint in the Heisenberg
pyrochlore antiferromagnet \eqr{eq:ground} entails
long-range correlations, but in the absence of
the polarization concept, the argument's form is diffuse,
and it was not possible to predict an explicit functional
form.  An interesting  heuristic argument was made there to justify the
empirical fact of the ``bowtie'' shape (see Fig.~\ref{fig:strucf})
taken by the diffuse scattering: in other words, 
that it takes a scaling form in terms of $q_x/q_\perp$.
\OMIT{Although a planar section of the diffuse
scattering does look like a ``bowtie'', in three-dimensional
reciprocal space the contours are surfaces of revolution.}

\subsection{Diffraction consequences}

\LATER{NEED introduxtory sentence}

\subsubsection{Spin structure factor}
\label{sec:spinstrucf-1}

There is a linear relationship
between the actual Ising spins and the ice arrows:
specifically, the map~\eqr{eq:Pol-alpha}, 
from spins to the the diamond-vertex polarization is actually
invertible:
   \begin{equation}
       t(\RR \pm \half \uu_m) = 4\uu_m \cdot \Pol(\RR), 
   \label{eq:Pol-spin}
   \end{equation}
where we take the $+$ or $-$ signs for even and odd
vertices $\RR$, respectively.
Thus it is not surprising that
the pseudodipolar correlations \eqr{eq:ice-dipolar}
of the latter imply similar $1/r^3$ correlations for the  former.
However, the coefficients relating these (namely,
the $\{ \uu _m \}$ vectors) are staggered in sign.

To see this, first Fourier transform \eqr{eq:Pol-spin}, using
the fact that 
    \begin{equation}
        \uu_{m,x}= \fourth e^{\half i\KK_{200}\cdot (\uu_m-\uu_4)}, 
    \label{eq:uuKK}
    \end{equation}
similarly $uu_{m,y}$ with $\KK_{020}$, etc.  When this is inserted
into the formula for $\tt(\kk)$ [similar to \eqr{eq:t-Fourier}], 
the staggering contained in the $\uu_m$
factors shifts the argument by the reciprocal space vector
$\KK_{200}$ so that $\tt(\KK_{200}+\qq) \propto P_x(\qq)$, etc.
Consequently, the singularities 
of the spin structure factor in reciprocal space 
are pseudo-dipolar in form, just like \eqr{eq:spindipolar}, but
occur at nonzero reciprocal lattice vectors $\KK$
rather than at $\qq=0$.

Let $\tt(\kk)$ be the Fourier transform of the spins.
The structure factor (i.e. the neutron diffraction intensity. 
modulo polarization factors) has pseudo-dipolar singularities
    \begin{equation}
          \Struc{\KK_{200}+\qq} \propto 
          \frac{q_\perp^2}{q_\parallel^2 + q_\perp^2}
   \label{eq:spindipolar}
    \end{equation}
Here $\KK_{200}\equiv [2\pi/a] (2,0,0)$
is a reciprocal lattice vector;
and $\qq_\parallel \equiv q_x$ and $q_\perp^2 \equiv q_y^2 + q_z^2$.
Near $\KK_{111}\equiv 2\pi(1,1,1)$, the same form holds with 
$\qq_\parallel$ and $\qq_\perp$ being the components of $\qq$
parallel and perpendicular to the $(111)$ direction.

The elastic constant $\kappa$ must be determined
from Monte Carlo simulations.
Yet, without knowing it, we  can still make the nontrivial prediction  that the 
diffuse scattering has quantitative the same strength near $\KK_{111}$
as it does near $\KK_{200}$. 
In  an isotropic Heisenberg model, 
these pseudodipolar singularities can be distinguished 
from true dipolar singularities
in neutron scattering experiments, 
since they are independent of  spin direction. 
They can be seen in only one spatial direction around each reciprocal
lattice point.

The functional form \eqr{eq:spindipolar}
has nodes  which we expect (correctly) are a consequence
of symmetry and therefore extend throughout reciprocal 
space, beyond the small-$\qq$ limit in which 
\eqr{eq:spindipolar} was derived.
For example, \eqr{eq:spindipolar} indicates that
   \begin{equation}
      \Struc{k_x,0,0} \equiv 0
   \label{eq:100zero}
   \end{equation}
for wavevectors anywhere along the entire (100) axis.
Similarly the structure factor is 
zero along the (111) axis 
near $\KK_{111}$ from \eqr{eq:spindipolar}
Since $\kk=(000)$ is the intersection of seven distinct axes,  
along each of which the structure factor must vanish, 
we deduce that the diffuse scattering vanishes near the origin as
   \begin{equation}
      \langle |t(\kk)|^2\rangle \propto
       (k_x^2  k_y^4 + {\text{5 permutations}})-6 k_x^2 k_y^2 k_z^2. 
   \end{equation}
Indeed, experiments and all simulations on pyrochlore
antiferromagnets find quite small diffraction 
throughout the first Brillouin zone.

\LATER{Citing Nagle~\cite{nagle}.}

\subsubsection{Derivation of nodal lines}

\def\tplane{{t_{\rm plane}}}

It is possible to confirm \eqr{eq:100zero}
without using coarse-graining, by sharpening
an argument made in Sec.~II D of Ref.~\onlinecite{moessner98b}.
(I have written the steps in detail, so it will be clear whether
they do or do not carry through for other lattices.)

Partition the pyrochlore lattice sites into (100) planes labeled by $x$, 
and define $\tplane(x) \equiv  \sum _{x_i=x} t_i$.
Then consider, for example, the tetrahedra with 
centers at $x=0$: each such tetrahedron includes a pair of spins 
with $x_i=-\fourth$ and another pair with $x_i=+\fourth$.
By the ground-state constraint \eqr{eq:ground} the sum of 
one pair is the negative of the sum of the other pair.  
Since every spin with $x_i=\pm \fourth$
belongs to exactly one such tetrahedron, 
this implies $\tplane(-\fourth)=-\tplane(\fourth)$;
and by Bravais lattice periodicity, 
$\tplane(\fourth+\half n)=(-1)^2 \tplane(\fourth)$.
Since all spins have $x_i=\fourth+\half n$ for some $n$, 
  \begin{equation}
      \tt(k_x,0,0) \propto \sum _n e^{-ik_x(\fourth+\half n)} 
      (-1)^n \tplane(\fourth)
  \end{equation}
which cancels, except for a possible Bragg peak when $k_x$ is a
multiple of $2\pi$.  
However, in the maximum entropy state that we
assumed in Sec.~\ref{eqc:FtotFT}, 
the average spin in one of these planes 
-- which is proportional to the mean polarization $P_x$ 
-- is zero, so $\tt(k_x,0,0)=0$ for all $k_x$.

A similar, but more complicated,  argument works for a $\{111\}$ direction.
Let $\xi_i\equiv \rr_i \cdot \frac {1}{\sqrt 3}(1,1,1)$ be the projection
of site $i$ on the $(111)$ axis, and redefine $\tplane(\xi)$ as the 
spin sum over the spin plane with $\xi_i=\xi$.  
Each tetrahedron with  $\xi=0$ includes
three spins in the $\xi=-1/4\sqrt{3}$ plane and one spin
in the $\xi=\sqrt{3}/4$ plane, and every spin in either plane
belongs to a unique tetrahedron; consequently 
$\tplane(-1/4\sqrt{3})=-
\tplane(+\sqrt{3}/4)$, 
even though one plane has three times as many spins as the other.
Furthermore these unequal planes are equally spaced, so
$\tplane(\sqrt{3}/4 + n/\sqrt{3}) = (-1)^n \tplane(\sqrt{3}/4)$
and the rest of the argument follows as before.

\subsubsection{Alternate derivation of nodal lines}

There is an alternate way to see that diffuse intensity must
be small near the zone center.  Let $\Pol^+(\kk)$
be Fourier transform of the discrete polarization 
on (say) even diamond sites, defined like \eqr{eq:L-Fourier}.
On the one hand, the  Fourier transform of the 
constraint \eqr{eq:ground} at {\it odd} vertices 
   \begin{equation}
       4 \G(-\kk)  \cdot \Pol(\kk) =0
   \label{eq:lattice-con-FT}
   \end{equation}
where 
   \begin{equation}
     \G(\kk) = \sum _{m=1}^4 e^{ i\kk\cdot \uu_m} \uu_m \approx
\fourth i \kk
   \end{equation}
at small wavevectors.
On the other hand, we can start from \eqr{eq:Pol-spin}
and take its Fourier transform -- this time, not taking
advantage of \eqr{eq:uuKK}; the result is
   \begin{equation}
      \tt(\kk) = 4 \G(\half \kk)\cdot  \tPol(\kk).
   \label{eq:Pol-spin-FT}
   \end{equation}
Near $\kk =0$, 
comparing \eqr{eq:Pol-spin-FT} and \eqr{eq:lattice-con-FT} 
and recalling that $\G(-\half \kk) \approx -\half \G(\kk)$, 
we conclude that $\tt(\kk)\approx 0$, as I asserted earlier.
Furthermore, whenever $\kk$ is along a $\{100\}$ or a $\{111\}$ symmetry axis, 
$\G(\kk)$ is parallel to that same axis by symmetry (for all $\kk$), 
hence structure factor \eqr{eq:100zero} is null all along those axes.

\subsection{Generalization to Heisenberg spins}
\label{sec:Heis-pol}

In the case of isotropic $n=3$ component spins, we can 
reiterate all the coarse-graining arguments of Sec.~\ref{sec:coarse}
in terms of the Heisenberg spin components.
Polarization components are defined as in \eqr{eq:Pol-alpha}
but now for each Cartesian spin component, so the polarization field is now
a {\it tensor} carrying not only space but also spin indices: 
$(\Pol_\alpha)_{\beta\mu}  \equiv  \sum _{i\in \alpha}
(\ss_i)_\beta (\uu_{m(i)})_\mu$.
As in the Ising case, it is easy to convince oneself that
the entropy density is maximum when $\Pol=0$. Hence we expect
that \eqr{eq:Fcoarse} remains valid, except that each $|\Pol(\rr)|^2$  
is now interpreted as a tensor norm (sum of squares of all
tensor elements). 
Finally, \eqr{eq:divergence} now becomes three equations, one for
each spin-space component. 

The basis of this paper 
-- and the signatures implied in the diffuse diffraction --
extends even to Heisenberg pyrochlore antiferromagnets
in a magnetic field $\Hfield$.  The replacement
$\HH \to \HH - \Hfield \cdot \sum_i \ss_i$
is equivalent to substituting 
  \begin{equation}
       \LL_\alpha \to \LL_\alpha - \Hfield/2\Jspin 
  \label{eq:L-shifted}
  \end{equation}
in \eqr{eq:Ham}. 
In the case of vector spins (but not Ising spins!), 
one achieves a ground state by
satisfying $\LL_\alpha=\Hfield/2\Jspin$ on every tetrahedron.
Simply replacing $\delta \ss_i\equiv \ss_i-\Hfield/8$
in the definition of $\Pol(\rr)$
(i.e subtracting off the mean spin expectation), 
provides a polarization appropriate 
to this ground state, which should exhibit the 
same sort of long-range correlations
(but with an $|\Hfield|$ dependence.)

\section{Lattice approximation}
\label{sec:lattice}

The continuum theory presented in  Sec.~\ref{sec:coarse}
can only predict the shape 
of singularities at special points in reciprocal space.
For a better comparison to experiments, 
a theory of the diffuse scattering 
throughout reciprocal space is desirable.
Of course, whereas the result \eqr{eq:ice-dipolar}
is universal across a class of models 
as enumerated in Sec.~\ref{sec:icemap}
and Sec.~\ref{sec:more-lattices}, 
the detailed shape of the scattering calculated here 
is specific to the pyrochlore lattice.

\OMIT{To handle the case of an external field, one 
would replace the $\delta$-functions
in \eqr{eq:maxlike}
by $\delta(\LL(\RR)-\Hfield)$.
But I don't trust the validity of the lattice derivation
with the Heisenberg model, so this is cut.}

The derivation depends on lattice Fourier transforms.
This unfortunately forces the introduction,  for this section, 
a new indexing $t(\RR+\half\uu_m)$ and $L(\RR)$
for the same objects $t_i$ and $L_\alpha$
defined previously, where $\RR$ designate the diamond sites;
the even sites $\RR$ form an fcc Bravais lattice and 
we let $N$ be the number of primitive unit cells.
Thus each pyrochlore site $\rr_i$ is written
$\RR+\half \uu_m$;  the surrounding odd
diamond vertices are at $\RR+\uu_m$.  
   \begin{equation}
     \tt_m(\kk)= \frac{1}{\sqrt N} 
                  \sum _\RR e^{-i\kk\cdot (\RR+\half \uu_m)} 
     t(\RR+\half \uu_m) .
   \label{eq:t-Fourier}
   \end{equation}
The definition \eqr{eq:L} is rewritten 
       $L(\RR)=\sum _{m=1}^4 t(\RR\pm\half \uu_m))$
taking $+$ and $-$ when $\RR$ is an even or odd diamond vertex,
respectively.  Thus the Fourier transform of \eqr{eq:L}, 
restricted to even or odd vertices  as indicated by the ``$\pm$'', is
   \begin{equation}
       \tilde{L}^{\pm} (\kk) =
      \frac{1}{\sqrt N} 
       \sum _{m=1}^4 e^{\pm \half i \kk\cdot \uu_m} \tt_m 
   \label{eq:L-Fourier} 
   \end{equation}
\OMIT{The phase factor used belongs to the actual
position of the spin, rather than to the center of the unit cell
associated with $\RR$.}

\subsection{Derivation of fluctuations}
\label{sec:lattice-deriv}

The simplest way to describe the approximation is
first to imagine a distribution of spins 
  \begin{equation}
          {\rm Prob}(\{ t_i \}) \propto 
      e^{-\sum _i \frac{t_i^2}{2 t_0^2}} \exp (-\betaeff \HH), 
  \label{eq:gaussianT}
  \end{equation}
where $\HH$ is given by \eqr{eq:Ham},
and $\betaeff$ ought to be the inverse temperature.
The Gaussian factor in \eqr{eq:maxlike}
may be viewed, in the Bayesian spirit of the maximum-likelihood
approach, as a trivial {\it a priori} independent distribution before we 
account for any spin interaction.  The second factor
enforces spin correlations; here $\HH(\{ t_i \})$ is the Hamiltonian
\eqr{eq:Ham} for Ising spins, except that now 
$t_i$ is allowed to  take any real value.

Adopting the limit $\betaeff \to \infty$, \eqr{eq:gaussianT} reduces to 
   \begin{equation}
          {\rm Prob}(\{ t_i \}) \propto 
      e^{-\sum _i \frac{t_i^2}{2 t_0^2}} 
   \prod_\alpha \delta(L_\alpha)
   \label{eq:maxlike}
   \end{equation}
The second factor in \eqr{eq:maxlike} imposes the ground state
constraints \eqr{eq:ground} around every tetrahedron, both even  and odd.

\subsubsection{Ising spin correlations}

Our goal is to evaluate $\langle t_i t_j \rangle$, and this is not hard
to do using the Fourier transform \eqr{eq:t-Fourier}.
We must first rewrite
$\prod_\alpha \delta(L_\alpha) 
\propto \prod_\kk \delta(L^+(\kk)) \delta(L^-(\kk))$
in \eqr{eq:maxlike}, and re-express this in terms of $\{ \tt_m(\kk) \}$.

Since different wavevectors decouple, it will be convenient to
view $\tt_m (\kk)$ as a complex 4-component vector.
Then the even and odd vertex constraints take the form 
of orthogonality conditions:
   \begin{equation}
      {\tilde L}^\pm(\kk) \equiv 
      \sum _{m=1}^4 e^{\pm \half i \kk\cdot \uu_m} \tt_m(\kk) 
      \equiv (E^\pm(\kk,\tt(\kk))=0
   \label{eq:orthog}
   \end{equation}
The coefficients are
   \begin{equation}
       E^\pm_m(\kk) \equiv e^{\mp\half i \kk \cdot \uu_m}
   \label{eq:E}
   \end{equation}
so $E^-_m \equiv (E^+_m)^*.$
The inner product is defined by
   \begin{equation}
         (A,B) \equiv \sum _{m=1}^4 A_m^* B_m.  
   \end{equation}

So, the distribution \eqr{eq:maxlike} 
can be rewritten in Fourier space as 
   \begin{equation}
       \prod_\kk \left[ e^{-\frac{1}{2 t_0^2} \sum_m^4 |\tt_m(\kk)|^2 }
       \delta((E^+(\kk),\tt) \delta((E^-(\kk),\tt) \right].
   \label{eq:spin-con-FT}
   \end{equation}
Thus, for each $\kk$ in \eqr{eq:spin-con-FT}, 
we now have a Gaussian distribution over a four-dimensional space, 
with two constraints reducing it to a two-dimensional subspace.  

\OMIT{[Technically, the space is eight
dimensional since the components $\tt_m$  are complex, but
the other four dimensions are ascribed to $\tt_m(-\kk) =
\tt_m(\kk)^*$, as is customary.]}
Our object is to obtain correlations of the form
    $\la \tt_l(-\kk) \tt_m (\kk) \ra$. 
The result will essentiallly be
the projector $\Pi$ into the subspace satisfying \eqr{eq:orthog}.
\LATER{Use slant font for $\Pi$?}
If we define the $4\times 2$ matrix $E$, the columns of which
are $E^+$ and $E^-$, 
then $4\times 4$ projection matrix for the (desired)
space orthogonal to $E^+$ and $E^-$ is  
$\Pi \equiv  I-E(E^\dagger E)^{-1} E^\dagger$.   
In fact
  \begin{eqnarray}
    (E^\dagger E) = 4 \pmatrix{ 1 & H\cr
                             H^* & 1} ,  \\
    (E^\dagger E)^{-1} = \frac{1}{4} (1-|H|^2)^{-1}
                      \pmatrix{ 1 & -H^*\cr
                               -H & 1}
   \label{eq:EEinverse}
   \end{eqnarray}
where the $\kk$ argument in $E$ and $H$ was suppressed; here 
  \begin{eqnarray}
      H(\kk) \equiv \frac{1}{4}  \sum _m e^{i\kk\cdot \uu_m} = 
        \cos \frac{k_x}{4} \cos \frac{k_y}{4} \cos \frac{k_z}{4} 
       \nonumber \\
         -i \sin \frac{k_x}{4} \sin \frac{k_y}{4} \sin \frac{k_z}{4} .
   \end{eqnarray}
[$H(-\kk)\equiv H(\kk)^*$,
$|H(\kk)|^2 = \frac{1}{4}(1+
        \cos \frac{k_y}{2} \cos \frac{k_z}{2} 
        + \cos \frac{k_x}{2}  \cos \frac{k_z}{2} 
        + \cos \frac{k_x}{2} \cos \frac{k_y}{2} )$.
The final result is 
   \begin{equation}
   \la \tt_l(-\kk) \tt_m (\kk) \ra =
       t_0^2 \left\{\delta_{lm}- [E(E^\dagger E)^{-1} E^\dagger]_{lm}\right\}
   \label{eq:tt-final}
   \end{equation}
where  $E$ and $(E^\dagger E)^{-1}$
are implicitly functions of $\kk$ and are defined by \eqr{eq:E}
and \eqr{eq:EEinverse}.

Since ${\rm Tr} \Pi =2$, we must have $t_0^2=2$ to satisfy
the normalization condition $\la t_i^2 \ra=1$.

\OMIT{If we want to be more formal, one can alternatively reach the
projection matrix by writing the two $\delta$ functions (for each $\kk$) as
$\delta(u_+) \delta(u_-)
=\int d\mu_+ d\mu_-  e^{i(\mu_+ u_+ + \mu_- u_-)}$; 
Switching the order of
integration over $\tt_m$ and over $\mu_+$, $\mu_-$ in the integrals, 
we can do the four-dimensional integrals over $\tt_m$ exactly using 
a shift of variable, and what remains is a two-dimensional 
Gaussian integral over the $\mu$ variables, in which the
$2\times 2$ matrix of coefficients is the same one as appears
in \eqr{eq:matrix}.}

\subsubsection{Spin structure factor}
\label{sec:spinstrucf-2}

The structure factor, as measured in neutron diffraction,
combines the four sublattice contributions.  
Write the Fourier  transform of all the spins as
$\tt(\kk) \equiv
     \frac{1}{\sqrt N} 
                  \sum_m \sum _\RR e^{-i\kk\cdot (\RR+\half \uu_m)} 
                  t(\RR+\half \uu_m) = (M,\tilde t(\kk))$, 
where we used the 4-vector inner product and 
    \begin{equation}
     M\equiv (1,1,1,1)
    \label{eq:Mcoeff}
    \end{equation}
The structure factor is then a projection of \eqr{eq:tt-final}
onto the $M$ vector, thus
   \begin{equation}
     \la |\tt(\kk)|^2\ra 
      = t_0^2 M^T [I- E(E^\dagger E)^{-1} E^\dagger] M .
   \label{eq:struc-result}
   \end{equation}

In the case of spin ice, the mapping of the Ising spins $\{ t_i \}$
to the real spins is different (it is staggered). 
In this case,  $M_{\beta m} = (\hat{\uu} _m)_\beta $.
The pinch-point singularities will appear, but at
different places (including the origin), because 
in this case the spins {\it are}  the local polarizations.
In this case, the  scattering around 
the origin is no longer suppressed.

In either the antiferromagnet or the spin-ice case,
there will be extinctions of the expected singularity
at points where $(E^{\pm}, M)$ happens to cancel.

\LATER{Work out the form of $|\Pol(\kk)|^2$ using a
similar technlogy, to verify the claims, 
connect to Sec. II, and extract the stiffness $\kappa$.}i

Considering the form of \eqr{eq:tt-final}, 
singularities in the spin fluctuations 
may (but do not necessarily) occur when the 
$2\times 2$ matrix $(E^\dagger E)$ is singular.
In view of \eqr{eq:EEinverse}, this occurs when
$|H(\kk)|^2=1$.  But that is precisely  the
defining condition for the fcc
reciprocal lattice vectors $\KK$.
[This was derived by a different route in
Sec.~\ref{sec:spinstrucf-1}.]

Indeed, $E^+(\KK_{200})=i(-1,-1,1,1)$ and 
$E^+(\KK_{111})=\frac{1+i}{\sqrt 2} (-1,1,1,1)$.
We see $E^+(\KK)$ is proportional
to a {\it real} four-vector.
[That generically happens at
a point in three-dimensional $\kk$-space, since there are 
three independent phase relationships to be satisfied 
among the components $\{ E^+_m \}$.]
Then $E^+$ is proportional to $E^-$, 
and the rank of $E(\KK)$ is reduced from two to one
at these points, confirming that $E(\KK)$ is singular.
\LATER{WORK OUT EXPLICITLY THE EXPANSION NEAR THESE POINTS.}

The special lines in reciprocal space on
which diffraction is zero, are those where
the 4-vector of coefficients \eqr{eq:Mcoeff},
relating the physical spin 
to the four sublattice
spins $\tt_m(\kk)$,  happens to be orthogonal
to both of the vectors in the null space of the 
$4\times 4$ matrix \eqr{eq:tt-final}.

\subsection{Finite temperature}

In the approximations of Sec.~\ref{sec:coarse}
and Sec.~\ref{sec:lattice-deriv}, 
where the tetrahedron constraint was imposed rigorously,
we were forced to project out the corresponding fluctuation modes.  
It is possible to extend the approximation so as to
permit small fluctuations of the tetrahedron magnetizations,
as must be excited at $T>0$, starting from
\eqr{eq:gaussianT}. Now, no projection
is required, since a large coefficient   $\betaeff$
tends to suppress those fluctuations, and
so this $\betaeff<\infty$ case is
actuallly more straightforward.
[This subsection is essentially 
a streamlining of Ref.~\onlinecite{yoshida},
as converted to my notations.]

The Hamiltonian in \eqr{eq:gaussianT} is a quadratic
form in the $t_i$'s. Fourier transforming\eqr{eq:Ham}, 
and putting it into the 4-vector notation with \eqr{eq:orthog}, we get
   \begin{equation}
      \HH =  \half J \sum _\kk \left[ |L^+(\kk)|^2 + |L^-(\kk)|^2 \right]
= \half J (\tt, E^\dagger E \tt)  .
   \end{equation}
Thus, \eqr{eq:gaussianT} becomes ${\rm Prob}(\{\tt_m\}) 
\propto \exp(-\half (\tt,\lambda \tt)$, where
   \begin{equation}
     \Lambda \equiv \frac{1}{t_0^2} I + \betaeff E^\dagger E.
   \end{equation}
and \eqr{eq:tt-final} gets replaced by
   \begin{equation}
     \la t_l(-\kk) t_m (\kk) \ra = (\Lambda^{-1})_{lm}.
   \label{eq:tt-thermal}
   \end{equation}
The matrix $E^\dagger E$ has rank two.
It can be seen that, as $\betaeff \to \infty$, 
\eqr{eq:tt-thermal} indeed reduces to the projector \eqr{eq:tt-final}.

The consequence of $\betaeff < \infty$ for correlations is that,  
in reciprocal space the pseudodipolar singularities 
\eqr{eq:spindipolar}, get rounded by the substitution 
   \begin{equation}
       \qq^2 \to \qq^2+\xi^{-2}
   \label{eq:screened}
   \end{equation}
in the denominator, 
where $\xi \sim \betaeff ^{1/2}$ is a a correlation length.
In real space, the power-law decays 
\eqr{eq:ice-dipolar} acquire an 
$\exp(-r/\xi)$ factor that cuts them off,
as noted in Ref.~\onlinecite{garanin}.

\subsection{Other analytic approximations}

Three prior treatments of the diffuse scattering in the
pyrochlore arrived
at a mathematical form more or less identical to \eqr{eq:gaussianT}, 
but with different formulas for the coefficients in these equations
as a function if temperature.
Thus, of course, their result is \eqr{eq:tt-thermal};
however, they did not note the pseudodipolar
(or, in the spin-ice case, literally dipolar) correlations
which are implicit in these formulas at the $T=0$ limit.



The diffuse scattering problem  was first addressed by 
Reimers~\cite{reimers92b} for Heisenberg, or in general 
$n$-component vector spins.)  That derivation is based on 
mean field theory (Ornstein-Zernike correlations), 
which ought to be valid in the critical regime $T\to \Tcmf$, 
where $\Tcmf=J/n$. 
However, that is invalid in the pyrochlore case, since
the real critical temperature is driven to zero.
Even so, the $\kk$ dependence of the result 
(eq.~(15) of Ref.~\onlinecite{reimers92b}), for the
limit $T\to \Tcmf$, is exactly of the form \eqr{eq:gaussianT}, 
with $1/t_0^2 = 3 (T/\Tcmf-1)$ and $\betaeff=1/T$. 


Canals and Garanin~\cite{garanin,canals-garanin}
considered the classical pyrochlore antiferromagnet with 
$n$-component unit spins, in the  large-$n$ limit which is 
tractable analytically.  This is, in effect, 
like the ``spherical model'' approximation for finite-$n$, 
in that a constraint $|\ss_i|=1$ on every spin
is replaced by one on all $N$ spins, $\sum_i |\ss_i|^2 = N$.
As they noted in Ref.~\onlinecite{garanin}, Sec. III, 
the $n=\infty$ limit is completely described by Gaussian 
approximations.
Their result for small temperatures can be reduced to 
\eqr{eq:gaussianT} with $1/t_0^2 = n/3$
and $\betaeff=1/T$;  this reduces to \eqr{eq:maxlike}
as $T\to 0$.

\LATER{CHECK Canals and G.}


Finally, Yoshida {\it et al}~\cite{yoshida} use an elaborate 
cluster-variational method to derive a sensible formula  for the
temperature-dependent diffuse scattering
for the {\it Ising} ($n=1$) pyrochlore antiferromagnet
(specifically, spin ice).
Their result, taking the lowest-order finite temperature correction, 
amounts to \eqr{eq:gaussianT} 
with  $t_0^2=2$ 
and $\betaeff J = \frac{3}{16} e^{2 J/T}$, 
which in the zero-temperature limit reduces to \eqr{eq:maxlike}. 
[Note that $J_{\rm eff}/3$ of Ref.~\onlinecite{yoshida} is my $J$.
My $E^{\pm m}$ are essentially linear combinations of the
4-vectors $c$ and $s$ of their eq.~(B-3).]

Until now, I have only mentioned approximations which 
involve the connectivity of the sites and which can incorporate
the long-range nature of the constraint.
Some other approximations, intended mainly to model the bulk
susceptibility, are based on a single isolated tetrahedron 
~\cite{harris95,moessner-berlinsky,garcia-huber}.
Apart from the restriction that this tetrahedron has 
zero net spin, 
these approaches necessarily miss the long-range constraint 
and hence give a poor picture of the long-range correlations. 

\subsection {Comparison to diffraction in 
experiment and simulation}

Several pyrochlore systems show the same characteristic 
diffuse scattering features:
(i) the entire first Brillouin zone has a very low intensity
 (ii) intensity is zero along $\{100\}$ abd  $\{111\}$
axes; (iii) there is a pinch-point singularity
of form \eqr{eq:spindipolar}
at $\KK_{200}$ and also $\KK_{111}$ reciprocal 
lattice vectors.
It should be noted that, since there are four  spin sites per
unit cell, the diffuse scattering 
is not periodic with the Brillouin zone.

Experimentally, such features were seen in an 
itinerant Laves phase (Ref.~\onlinecite{ballou}, Fig.~3);
in the pyrochlore CsNiCrF$_6$''
(Ref.~\onlinecite{harris97a}, Fig.~4);
and most recently in the spinel
ZnCrO$_4$ (Ref.~\onlinecite{lee02}, Fig.~3(a,b)), 
in the higher-temperature regime above  
a structural transition.
In simulations, such patterns appeared
in Fig.~2 of Ref.~\onlinecite{zinhar95},
Fig.~4 of ~Ref.~\onlinecite{harris97a}, 
and in Ref.~\onlinecite{moessner98b}.
For comparison, spin-1/2 results
from exact diaagonalization are
shown in Ref.~\onlinecite{canals98}, Fig.~4;
they are qualitatively similar, but less sharp.

Images of analytic large-$n$ calculations
of Garanin and Canals can also be compared:
Ref.~\onlinecite{canals-garanin}, Fig.~4
(which is the \kag system)
and  Ref.~\onlinecite{Can00}, Fig.~6.

\LATER{Add Yoshida to this list.}

Ref.~\onlinecite{moessner-kagice}, in their
Fig.~5, plot diffuse scattering from 
simulation of a two-dimensional spin problem 
equivalent to the honeycomb dimer covering.
(This is a plane of spins in the
``\kag-ice'' phase, whereby an external field applied to
a ``spin-ice'' pyrochlore system causes $\{111\}$ planes to decouple). 
The pinch-points (called ``bowties'' by those authors) are 
prominently visible, which are diagnostic of pseudodipolar correlations 
on real space.

The structure factor \eqr{eq:spindipolar} 
has a local maximum at $q+x=0$, if
we vary $q_x$ along a line $\qq_\perp={\rm const}$, 
offset slightly from the $q_x$-axis.
The same behavior is found around $\KK_{111}$ and, 
of course, all other symmery-equivalent reciprocal lattice vectors:
the structure factor has maxima in the plane perpendicular to 
the radial direction in reciprocal space.
The union of these planar facets forms the same shape (a cuboctahedron)
as the boundary of the FCC lattice's first Brillouin boundary, 
but doubled in all three directions.
Indeed, in a plane of reciprocal space from a  single crystal
of Y$_{1-x}$Sc$_x$Mn$_2$, the diffuse scattering 
is concentrated near the {\it lines} where this
plane cuts the Brillouin zone faces \cite{ballou}.

Powder diffraction data from pyrochlores showed 
a characteristic maximum 
  ~\cite{reimers92b}
which was seen experimentally~\cite{harris94}, 
and in simulation~\cite{reimers92a}.
\OMIT{Note that $a\approx 1.02$ nm in the pyrochlore lattice.)}
This is consistent with the fact that the (doubled)
Brillouin zone boundary (where diffraction is 
maximum along any ray in reciprocal space) is roughly a sphere.
(Note that powder averaging, in the vicinity of $\KK_{200}$, amounts to 
integrating \eqr{eq:spindipolar} over $q_y$ and $q_z$, which yields 
a weakly cusped function ${\rm Const}- 2 \pi q_x^2 \ln q_x$.)

\section{Discussion}
\label{sec:disc}

In summary, it was found that a polarization can be
defined in pyrochlore antiferromagnets which 
-- in a ground state -- exactly satisfies a 
divergence condition. The coarse-grained version 
of this is analogous to an order parameter, 
in being the natural variable to describe
large-scale properties.  
The correlations were found (Sec.~\ref{sec:coarse})
to have a pseudo-dipolar form  which -- since this is a 
pure power law -- implies an infinite correlation length.
These behaviors were repeated in a lattice-based derivation
(Sec.~\ref{sec:lattice}.  
In the rest of this section, I discuss other problems 
to which these findings or approaches could be extended 
or related.

\subsection{Implications for diffraction experiments}

\label{sec:neutrons}

The argument of this paper suggests that
the analytis of diffraction experiments
ought to focus more on the characteristic features,
such as pinch points, identified in 
Sec.~\ref{sec:spinstrucf-1} and Sec.~\ref{sec:spinstrucf-2}.
Deviations from the predictions at those places
are sensitive measures of the extent to
which the tetrahedron constraint is violated
in the actual ensemble.   Thus, it is suggested
to analyze these experiments so as to extract
the correlation length $\xi$, and to check how
well the diffraction is suppressed along the 
predicted nodal lines.
[A correlation length
was extracted in Ref.~\onlinecite{zinhar95}
from simulations, however in this case it was
actually the finite size cutoff.]

Deviations in the overall pattern from the shape
predicted in Sec.~\ref{sec:spinstrucf-2}
(which are consistent with simulations~\cite{moessner-dipolar})
are likely to reflect additional
terms in the Hamiltonian, as in Ref.~\onlinecite{lee02}.

Ref.~\onlinecite{lee02} have fitted the diffuse intensity as
the Fourier transform, not of a single tetrahedron, but a single loop
of 6 spins.
However, contrary to the speculation in that paper, this does 
not necessarily indicate a physical state built from such hexagons.
To explain this, I will outline an alternative, equally
systematic way of fitting the diffuse diffraction data.

The constraint \eqr{eq:Ham} implies a similar constraint on
the matrix of correlation functions.  Then one can express any
valid correlation function  using a 
basis of linearly independent, orthogonal functions in real space satisfying
this constraint, in the spirit of e.g. spherical harmonics.  
The first of these terms is the same correlation that derives 
from the ring of six. The form observed~\cite{lee02} is, 
from this viewpoint, the simplest possible shape, as one
might expect at higher temperatures when all the other
terms are damped out.  To produce a pinch point, an infinite
number of such functions would be needed, corresponding to
a large spatial extent.

\subsection{Dynamics }

It is well known that violations of the tetrahedron 
constraint (due to disorder, or thermal excitation)
map to electric charges (in the language where $\Pol(\rr)$
is an ``electric field.''
A defect in which three arrows point
outwards has a ``charge'' of $Q=+1$, or $Q=-1$ if three point
inwards~\cite{FN-Gausslaw}; in general, $Q=L_\alpha/2$.
These ``charges'' can only be created in opposite
pairs;  such a pair feels an (entropic)
effective interaction which behaves (at coarse-grained
distances) exactly like a Coulomb interaction.  
At $T>0$, the interaction would be screened in the usual
fashion, and the screening length is the correlation 
length mentioned in Subsec.~\ref{sec:neutrons}.

Apart from these defects, the total (classical) spin is constrained to
be exactly zero. [The total magnetization is
$\half \sum_\alpha{\LL_\alpha} =0$ by \eqr{eq:ground}.]
Hence, the defects are central to any theory of the
magnetic relaxation,  as observed by inelastic neutron diffraction, 
or in AC susceptibility 
measurements~\cite{snyder03,snyder-dilution}.

To sharpen this point, note that within the ground states,
there is {\it no} local move that produces another valid state:
an entire loop must be updated at once~\cite{stillinger}.
But the movement of a charge along a loop leaves behind the
same change and thus implements this nonlocal ``flip'' operation.

The interpretation of relaxation experiments  ought to be cast in 
terms of the diffusion  and recombination of pseudo ``electric charges".
The relaxation rate of the real magnetization is proportional
to the drift mobility of the ``charges''.  The theory of their
behavior is isomorphic to an intrinsic semiconductor;
the cost a $Q=\pm1$ defect (in an Ising model) is $2J$, so
$4J$ plays the role of the bandgap.

Nonmagnetic impurity sites~\cite{snyder-dilution} 
act like impurity levels in a semiconductor
(exactly at midgap).  In the Ising model, a tetrahedron with 
one missing site has a ground state with $L_\alpha=\pm 1$,
corresponding to $Q=\pm 1/2$: it is as if a quenched charge
of $-1/2$ has been placed there, with the possibility of
binding a $+1$ charge to it.
In a Heisenberg model, such a tetrahedron often
still satisfies $\LL_\alpha=0$ and the behavior is
more subtle~\cite{henley-HFM00}.

We could also make predictions for the dynamic neutron 
structure factor.  The polarization $\Pol(\rr)$ is a conserved
quantity, so (in a classical model) it must diffuse.
Thus
  \begin{equation}
      \frac{\partial \Pol}{\partial t} = - \nabla \cdot \JJ_\Pol
      = \Gamma \nabla^2 \Pol
  \label{eq:diffusion}
  \end{equation}
Hence, near a pinch point $\KK$, this [classical] dynamics
implies vanishing widths in the dynamic structure  factor 
${\cal S}(\KK+\qq,\omega) \sim \Gamma \qq^2$.  
Dynamical conclusions were derived by Moessner and Chalker
directly from an equation of motion in terms of $\LL_\alpha$.
[The consequence in a quantum model appears to be
a gapless excitation with linear dispersion~\cite{hermele}.]

\subsection{Other models}

\subsubsection{Other three-dimensional lattices}
\label{sec:more-lattices}

The polarization construction can be generalized to any model 
in which (i)  ``spins'' sit on the edges of a bipartite 
graph, and (ii) around every vertex, the sums of the ``spins''
is the same;  here ``spin'' degree might be structural 
as well as magnetic.  

In three dimensions, the best 
examples (apart from the pyrochlore lattice) are 
antiferromagnets on the garnet lattice~\cite{petrenko} 
or dimer coverings on the simple cubic, bcc, and diamond
lattices.
Dimer coverings of the diamond lattice
might realized by the ice model
in  which ions of difference valences sit on the even
and odd diamond sites; they also correspond 
to the ground states of an Ising pyrochlore in
an external field.
This last system is most
plausibly realized by a  1/4-occupancy lattice gas,
with nearest-neighbor repulsion,
representing a charge order problem.
Alongside the dimer models are vertex models
on the simple cubic or triangular lattice,
in which each site has three inwards and three
outwards arrows~\cite{balents,hermele}. 

\subsubsection{Two-dimensional models}

In two dimensions,
constraints such as \eqr{eq:ground} are encountered in several models, 
notably in two-dimensional ice (= six-vertex model), the
triangular Ising antiferromagnet ground state, 
the square lattice dimer covering, and especially the 
\kag Heisenberg antiferromagnet~\cite{kagome,huse,kondev}.
To compare the last of these to the pyrochlore Ising ground states, it is
fairest to consider the ground states of the three-state Potts model 
antiferromagnet on the \kag lattice.
In that system, as in the pyrochlore, one can predict the structure factor
$\Struc{\kk}=0$ along the lines through the origin and
its first star of reciprocal lattice vectors; and here too
the scattering tends
to concentrate along a surface which is the Brillouin zone boundary, 
scaled up by a factor of 2. 

Following the analogy to the pyrochlore, one would guess 
the scattering has a smooth maximum at the zone corner 
$\QQ\equiv \{ {2\over 3}, {2\over 3}\}$ 
type points, but in this respect point the behavior differs from $d=3$.
To understand that, consider that \eqr{eq:divergence} is
satisfied formally by writing
   \begin{equation}
   \Pol(\rr) = \nabla \times \hh(\rr) 
   \end{equation}
where $\hh(\rr)$ is a ``vector potential''.
In $d=2$, there is no gauge freedom:
for a given configuration $\{\Pol(\rr)\}$,
$\hh(\rr)$ is {\it uniquely} determined (apart from a constant),
and can be visualized as 
parametrizing a (rough) interface in a $2+1$ dimensional 
space~\cite{blote}.
Following standard ``Kosterlitz-Thouless'' (also known as
``Coulomb-gas'') prescriptions, spin operators
have a component which is a periodic function of the
local $\hh(\rr)$. This implies  correlations
with a power-law proportional to $1/\kappa$, 
and the structure factor 
must have a power-law cusp at $\QQ$
(which I will call a ``zone corner singularity'')
reflecting this quasi-long-range order~\cite{huse}.
Ref.~\onlinecite{moessner-kagice} have noticed 
these ``zone-corner singularities'' in simulations;
they also  appear when one
directly measures the fluctuations $\la |h(\qq)|^2 \ra$
of a discretely defined  height field~\cite{henley-height,zeng}.

In such ``height models''  it is also possible that
the free energy favors a state with bounded fluctuations
of $\hh(\rr)$, corresponding to long-range order of the 
spins.~\cite{huse,zeng}. 
In the rare cases of a height
model in $d=3$ (e.g. three-state Potts antiferromagnet
on the simple cubic lattice), or in the ground state
of a $d=2$ quantum system, 
this ``locking'' behavior is always expected, except in 
some quantum models which contain nontrivial Berry phases. 

Ref.~\onlinecite{moessner-kagice} have calculated
the asymptotic correlation function for 
the dimer coverings of a triangular lattice, 
a system which is realized in the ``spin-ice''
class of pyrochlore in a magnetic field
oriented along $\la 111 \ra$: see their
eqs.~(4.3)-(4.4). 
It should be noted that two-dimensional dimer models
(which are solvable by the methods used for the
Ising model, i.e. free fermions),  have
the peculiarity that 
the correlations arising from the
``height'' field $\hh(\rr)$ have
exactly the same decay ($1/r^2)$ as
the pseudodipolar terms;
the contributions can be distinguished 
because the first kind of correlation does not 
depend on the orientation of the vector $\rr$
between sites, while the pseudo-dipolar kind does.

\LATER{CHECK THAT Ref.~\onlinecite{henley-height}
do both mention zone-corner singularities.}

\subsubsection{Quantum models}

For the $S=1/2$ case, there is 
believed to be no spin order~\cite{canals98}.
The theory of this paper is not literally applicable
to small-$S$ auantum systems, since no wavefunction is
possible in which every tetrahedron is simultaneously
a singlet.  Nevertheless, it is claimed~\cite{canals-garanin}
that the diffraction from exact diagonalizations 
of the spin-1/2 case agrees well with formulas such as
\eqr{eq:struc-result}.

\OMIT{Critique Fisher et al for using 6-loop update:
is it really ergodic? Also, cite Newman et al
algorithm for ice.}

A pyrochlore Ising antiferromagnet, made into a
quantum model by a small transverse ring exchange, 
has also defined the same coarse-grained field
$\Pol(\rr)$, which in their theory is called
an ``electric field''~\cite{hermele}.  
Out of their quantum-mechanical
variables conjugate to $\Pol(\rr)$ they construct
a ``magnetic  field''; the resulting $U(1)$ gauge
theory has gapless modes with $|\kk|$ dispersion
analogous to light waves, and correspondingly
there are power law correlations (though with
a different power law.)   Ref.~\onlinecite{huse-dimer}
also studied a model with a polarization 
(they call $\Pol(\rr)$ the ``magnetic'' field)
as a means to constructing quantum models with
no long range order and fractionalized excitations.

Recently, interesting phenomena have been observed in
certain (electron) {\it conductors} containing a 
pyrochlore sublattice~\cite{fulde}, which are
speculated to be related to the frustration of
this lattice.
In particular,
heavy fermion behavior is seen~\cite{fulde} 
in the spinel $\rm LiV_2 O_4$, 
and unsual ferromagnetic behavior in pyrochlore $\rm Nd_2Mo_2O_7$
is ascribed  to a Berry phase acquired by
the fermions in a spin background~\cite{nagaosa}.
Perhaps the coarse-grained polarization
field can help in modeling the long-wavelength behavior
of these systems.

There is also a speculation~\cite{ihm}
that the low-temperature state
of ice itself (neglecting the dipole couplings beyond the nearest
neighbor!) is dominated by proton tunneling with nontrivial
Berry phases.
Even though this model is (probably) not relevant to real ice, 
its exotic ground state is of interest in its own right
and the $\Pol(\rr)$ field is likely to enter its decription.

\subsection{Diffraction singularities due to constraints}

Local constraints produce diffraction singularities 
in other systems, 
specifically in the so-called ``transition state"
of certain metal alloys~\cite{deridder}.
In an FCC lattice, let $ (\RR) =\pm 1$ represent two
different chemical species. 
To model a state which has strong short-range order,
assume a constraint resembling \eqr{eq:ground}, 
  \begin{equation}
        \sum _m s(\RR+\uu_m) =0, 
  \label{eq:tr-st-con}
  \end{equation}
where $\{ \RR+ \uu_m \}$ are the 12 nearest-neighbor sites.
After Fourier transforming, we obtain
  \begin{equation}
      F(\kk) {\tilde s}(\kk) =0, 
  \label{eq:tr-states}
  \end{equation}
where $F(\kk) \equiv \sum _m \exp (i\kk\cdot \uu_m)$. 
It follows  from \eqr{eq:tr-states} this that the diffuse scattering 
$\langle |{\tilde s}(\kk)|^2 \rangle$ 
is {\it zero everywhere} in reciprocal space, 
except that it {\it diverges}~\cite{deridder}
along the two-dimensional surfaces defined by $F(\kk)=0$.
When these surfaces intersect the Ewald plane of an 
electron diffraction experiment, they produce
well-known arcs observed in fcc metal alloys near ordering transitions 
~\cite{deridder}. 
Similar behavior is seen in the $\rm Na_{22}Ba_{14}CaN_6$ structure,  
a triangular arrangement of rods each having an Ising degree of
freedom with ``antiferromagnetic'' correlations~\cite{tri-rods}.
Arcs are also seen in quasicrystals, where they are
ascribed to the constraints of tiling space~\cite{Ne95},

Eq.~\eqr{eq:tr-states} is a sharper singularity
than is found for the pyrochlore lattice in this paper.
The fundamental reason is that  the ``spins'' in
\eqr{eq:tr-st-con} are on a 
primitive Bravais (FCC) lattice, so in a lattice
of $N$ cells  there are $N$ constraints and the same
number of variables.
In contrast, in the pyrochlore problem analyzed in the present paper,
there are four sites per primitive cell
and only two constraints.
Consequently \eqr{eq:tr-states} gets replaced by 
the $4\times 2$ matrix equation \eqr{eq:orthog}.
Just  as \eqr{eq:tr-states} has its singular surfaces at $\kk$
values where the number of constraints, so \eqr{eq:orthog}
has its pseudodipolar singularities
at the points (reciprocal lattice vectors) at which the 
two equations are linearly dependent and reduce to one equation.

The above constraint-counting arguments have been 
phrased so as to make clear how they might be 
adapted to other systems.  For example, the triangular
Ising antiferromagnet (in zero field) is a highly frustrated system
on a Bravais lattice, so one might naively expect 
stronger kinds of singularity in its diffuse scattering. 
However, in that case there is no equality constraint but 
instead $\sum _{i\in \alpha}t_i =\pm 1$, so the whole
approach breaks down.

\acknowledgments

\OMIT {M. Kvale for unpublished collaboration and discussions on 
the pyrochlore antiferromagnet.}
I thank R. Ballou and C. Broholm  for sharing data before
publication, 
J. F. Nagle and F. Stillinger for information on the ice model,
as well as O. Tchernyshyov, R. Moessner, and S. Sondhi for discussions.
This work was supported by the National Science Foundation 
(NSF) grant no. DMR-0240953.
Part of it was completed at the Kavli
Institute for Theoretical Physics, supported 
by NSF grant PHY99-0794.

\OMIT{
I really won't make a point of powder data. 
It all looks the same -- a hump at about the right place, followed
by a valley. (Indeed, when Reimers back-transformed the scattering to
obtain radial correlations, there were some big differences 
among his various papers of 1991 -- in one case he concluded nearest
neighbor ferromagnetic interactions, which seems bizarre. 
That's just the data which most closely resembles the
expected hump, Fig. 9 on p. 5686 of [greeet91].
It is for $\rm Tb_2 Mo_2 O_7$. 
Furthermore both Tb and Mo have moments, and he claims that 
the dominant coupling is AFM Tb-Mo; therefore, this isn't 
a clean example!}

\OMIT{The powder data from $\rm CsMnFeF_6$ in W. Kurtz et al, 
in Sol. State. Comm. 18, 1479 (1976), seems to show such a hump. 
But it is suspicious: Laurent L\'evy told me (1987) that 
$\rm CsNiFeF_6$ and $\rm CsMnFeF_6$  have strong FM tendencies. 
Fig. 4.9 in Liebmann  quotes {\it single-crystal} data 
from the same group
(M. Steiner et al, J. Phys. Soc. Jpn. Suppl. 52, 173 (1983) --
I have this somewhere]. It clearly doesn't look much like
the expectation from nearest-neighbors (e.g. Zinkin \& Harris;
a cruder version from MC is shown in 
Liebmann's Fig. 4.12; including weakish nnn interactions was enough 
to radically change it to look like the data (his Fig. 4.13). 
}

\begin{thebibliography}{99}
\bibitem{FN-Gausslaw}
With this normalization of
charge, each ice-arrow must carry a discrete flux of $1/2$.
By the same rule,the maximum-polarization state 
(which by \eqr{eq:Pol-alpha} is $\Pol=(1,0,0)$) carries a flux
per unit area of $2/a^2$ across a $(100)$ plane.  
Hence, to satisfy Gauss's law, we must use a coarse-grained
polarization  $2 P(\rr)/a^2$.
\end{thebibliography}
\end{document}